\documentclass[fleqn,10pt]{wlscirep}
\usepackage{amssymb}
\usepackage{amsmath}
\usepackage{graphicx}
\usepackage{epsfig}

\title{Maximal distant entanglement in Kitaev tube}
\author[1]{P. Wang}
\author[1]{S. Lin}
\author[1,2]{G. Zhang}
\author[1,*]{Z. Song}
\affil[1]{School of Physics, Nankai University, Tianjin 300071, China}
\affil[2]{College of Physics and Materials Science, Tianjin Normal University, Tianjin 300387, China}
\affil[*]{songtc@nankai.edu.cn}
\begin{abstract}
We study the Kitaev model on a finite-size square lattice with periodic
boundary conditions in one direction and open boundary conditions in the
other. Based on the fact that the Majorana representation of Kitaev model is
equivalent to a brick wall model under the condition $t=\Delta =\mu
$, this system is shown to support perfect Majorana bound states which is in
strong localization limit. By introducing edge-mode fermionic operator and
pseudo-spin representation, we find that such edge modes are always
associated with maximal entanglement between two edges of the tube,
which is independent of the size of the system.
\end{abstract}
\maketitle

\begin{document}

\newpage

\section*{Introduction}

Topological materials have become the focus of intense research in the last
years \cite{Hasan,XLQ,CKC,HMW}, since they not only exhibit new physical
phenomena with potential technological applications, but also provide a
fertile ground for the discovery of fermionic particles and phenomena
predicted in high-energy physics, including Majorana \cite%
{LF,RML,VM,SNP,YO,NR}, Dirac \cite{AHCN,ZKL,ZKL2,JAS,ZW,JX,SMY} and Weyl
fermions \cite{MH,SMH,BQL,BQL2,CS,XW,HW,SYX,SYX2}.\ These concepts relate to
Majorana edge modes. A gapful phase can be topologically non-trivial,
commonly referred to as topological insulators and superconductors, the band
structure of which is characterized by nontrivial topology. The number of
Majorana edge modes is determined by bulk topological invariant. In general,
edge states are the eigenstates of Hamiltonian that are exponentially
localized at the boundary of the system. A particularly important concept is
the bulk-edge correspondence, which links the nontrivial topological
invariant in the bulk to the localized edge modes. On the other hand,
Majorana edge modes have been actively pursued in condensed matter physics
\cite{AJ,BCWJ,STD,LM,ESR,DSS,SM} since spatially separated Majorana fermions
lead to degenerate ground states, which encode qubits immune to local
dechoerence \cite{NC}. There have been theoretical proposals for detecting
Majorana fermions in 2D semiconductor heterostructures \cite{SJD,AJ2},
topological insulator-superconductor proximity\textbf{\ }\cite%
{LF,QXL,CSB,LKT,AAR}, 1D spin-orbit-coupled quantum wires \cite%
{RML,YO,AJ3,LRM2,STD2,PAC,RD,PE,DSS2,LJ} and cold atom systems \cite%
{SM,ZC,TS,LXJ,JL,DS}. Experimentally, it is claimed that indirect signatures
of Majorana fermions in topological superconductors have been observed \cite%
{VM,SNP,RLP,DMT,DA,FADK,CHOH,XJP,SHH,WMX,WZF}\textbf{.} So far, the
thoretically predicted Majorana bound state in literatures requires the
system in thermodynamic limit. An interesting question is whether there
exists the Majorana bound state in a small sized system, or the topological
feature is a prerequisite for Majorana bound state.\ The existence of such a
type of edge mode would indicate that the bulk topology is not necessary to
the spatially separated Majorana fermions and\ may provide an alternative
way to detect and utilize\ Majorana fermions.

In this paper, we study the Majorana edge modes in the Kitaev model on a
square lattice based on analytical solutions. In contrast to previous
studies based on open boundary conditions in two directions, we focus on a
finite-length cylindrical lattice. We show that the Majorana representation
of Kitaev model is related to a brick wall model, based on which this model
in a finite-length cylindrical geometry supports the perfect Majorana bound
states \textbf{under the condition} $t=\Delta =\mu $. The perfect Majorana
bound state is in the strong\ localization limit. This Majorana zero mode
has two notable features: (i) The edge-mode states exhibit maximal
entanglement between the two edges of the cylinder; (ii) By introducing
edge-mode pesudospin operators, we find that the edge mode relates to a
conserved observable. Remarkably, the expectation values of two types of
pseudospins for eigenstates indicate the coexistence of both bosonic and
fermionic excitations. And the eigenstates also possess maximal entanglement
about the bosonic and fermionic modes. These results provide a way to detect
the Majorana bound states in $p$-wave superconductors.

\section{Model}

\label{Model} We consider the Kitaev model on a square lattice which is
employed to depict $2$D $p$-wave superconductors. The Hamiltonian of the
tight-binding model on a square lattice takes the following forma%
\begin{eqnarray}
H &=&-t\sum_{\mathbf{r,a}}c_{\mathbf{r}}^{\dagger }c_{\mathbf{r}+\mathbf{a}%
}-\Delta \sum_{\mathbf{r},\mathbf{a}}c_{\mathbf{r}}c_{\mathbf{r}+\mathbf{a}%
}+h.c.  \notag \\
&&+\mu \sum_{\mathbf{r}}\left( 2c_{\mathbf{r}}^{\dagger }c_{\mathbf{r}%
}-1\right) ,
\end{eqnarray}%
where $\mathbf{r}$ is the coordinates of lattice sites and $c_{\mathbf{r}}$
is the fermion annihilation operators at site $\mathbf{r}$. Vectors $\mathbf{%
a}=a\mathbf{i},$ $a\mathbf{j},$ are the lattice vectors in the $x$ and $y$
directions with unitary vectors $\mathbf{i}$\ and $\mathbf{j}$. The hopping
between (pair operator of) neighboring sites is described by the hopping
amplitude $t$ (the real order parameter $\Delta $). The last term gives the
chemical potential. Imposing boundary conditions on both directions, the
Hamiltonian can be exactly diagonalized. The Kitaev model on a honeycomb
lattice and chain provides well-known examples of systems with such a
bulk-boundary correspondence \cite{AK,GB,DHL,KPS,GK,GK2,GK3}. It is well
known that a sufficient long chain has Majorana modes at its two ends \cite%
{AYK}. A number of experimental realizations of such models have found
evidence for such Majorana modes \cite{VM,LPR,DA,FADK,BA}.\textbf{\ }In
contrast to previous studies based on system in thermodynamic limit, we
focus on the Kitaev model on a finite lattice system. This is motivated by
the desire to get a clear physical picture of the egde mode via the
investigation of a small system. We first study the present model from the
description in terms of Majorana fermions.

We introduce Majorana fermion operators%
\begin{equation}
a_{\mathbf{r}}=c_{\mathbf{r}}^{\dagger }+c_{\mathbf{r}},b_{\mathbf{r}%
}=-i\left( c_{\mathbf{r}}^{\dagger }-c_{\mathbf{r}}\right) ,
\end{equation}%
which satisfy the relations%
\begin{eqnarray}
\left\{ a_{\mathbf{r}},a_{\mathbf{r}^{\prime }}\right\} &=&2\delta _{\mathbf{%
r},\mathbf{r}^{\prime }},\left\{ b_{\mathbf{r}},b_{\mathbf{r}^{\prime
}}\right\} =2\delta _{\mathbf{r},\mathbf{r}^{\prime }},  \notag \\
\left\{ a_{\mathbf{r}},b_{\mathbf{r}^{\prime }}\right\} &=&0,a_{\mathbf{r}%
}^{2}=b_{\mathbf{r}}^{2}=1.
\end{eqnarray}%
Then the Majorana representation of the Hamiltonian is%
\begin{eqnarray}
&&H=-\frac{1}{4}\sum_{\mathbf{r}}[i(t+\Delta )\sum_{\mathbf{a}}a_{\mathbf{r}%
}b_{\mathbf{r+a}}  \notag \\
&&+i(t-\Delta )\sum_{\mathbf{a}}b_{\mathbf{r}}a_{\mathbf{r+a}}+i2\mu a_{%
\mathbf{r}}b_{\mathbf{r}}+h.c.].
\end{eqnarray}%
It represents a dimerized brick wall lattice (or honeycomb lattice) with
extra hopping term $b_{\mathbf{r}}a_{\mathbf{r+a}}$.
\begin{figure}[tbph]
\centering
\includegraphics[bb=335 188 1154 680, width=15 cm, clip]{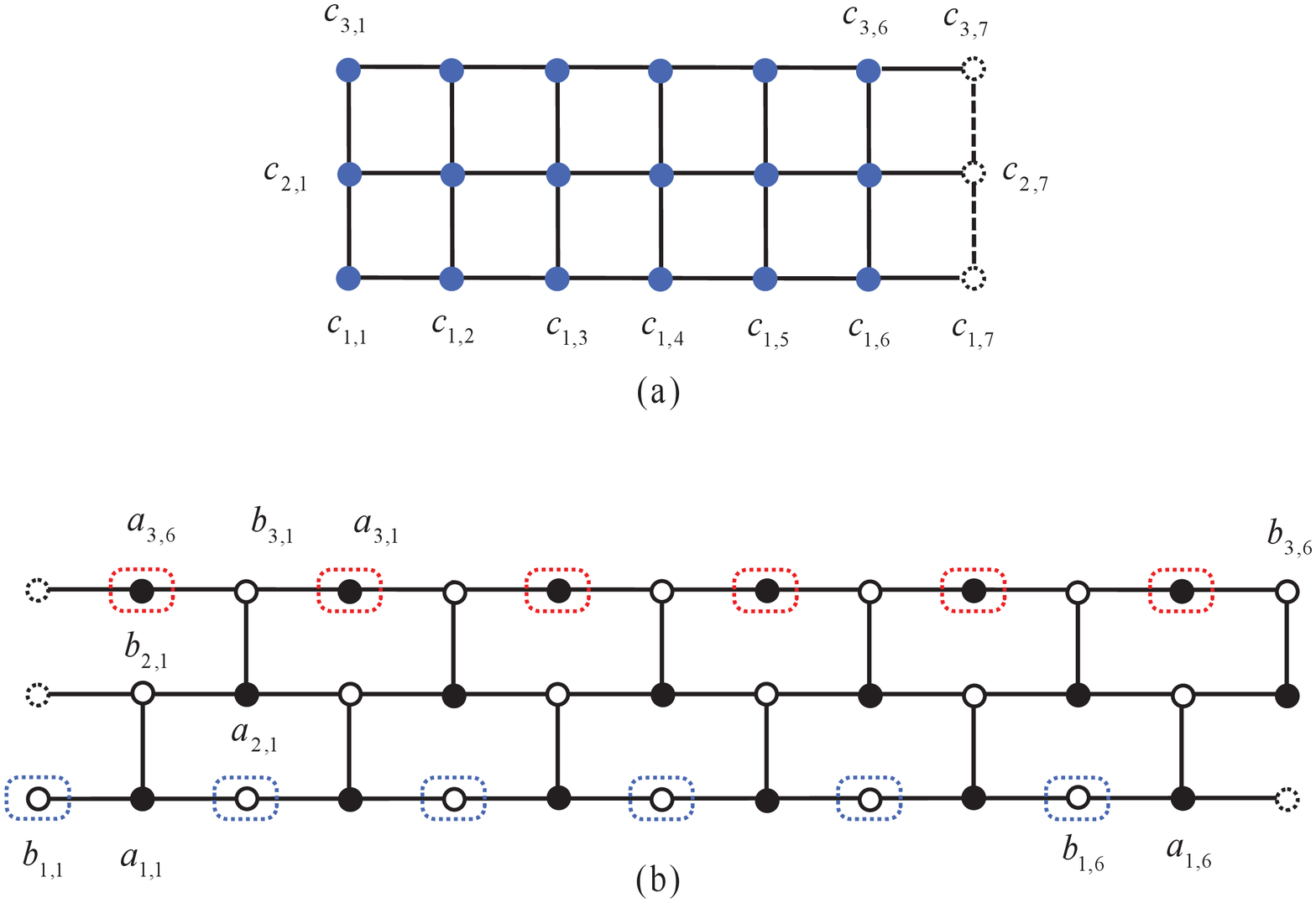}
\caption{(Color online) Schematic picture of the Kitaev model on a square
lattice and its corresponding Majorana fermion system. (a) A $3\times 6$
square lattice with periodic boundary condition in horizontal direction and
open boundary in vertical direction. (b) The corresponding Majorana system
which is a brick wall lattice with the same boundary conditions in lattice
(a). Fermions $c_{i,j}$ (blue circle) in (a) are decomposed into two
Majorana fermions $a_{i,j}$ and $b_{i,j}$ (white and black circles,
respectively) in (b). Majorana edge states for $a$ and $b$ are indicated by
blue and red dotted circles, respectively, which are perfectly localized at
the two edges of the cylinder.}
\label{fig1}
\end{figure}

\section{Majorana edge modes}

\label{Majorana edge modes} Let us consider a simple case to show that
Majorana modes can appear on some edges.\ Taking $t=\Delta =\mu $ the
Hamiltonian reduces to%
\begin{equation}
H_{BW}=-\frac{t}{2}\sum_{\mathbf{r}}(ia_{\mathbf{r}}\sum_{\mathbf{a}}b_{%
\mathbf{r+a}}+ia_{\mathbf{r}}b_{\mathbf{r}}+h.c.).
\end{equation}%
which corresponds to the original Kitaev model

\begin{eqnarray}
H_{BW} &=&-t\sum_{\mathbf{r,a}}\left( c_{\mathbf{r}}^{\dagger }c_{\mathbf{r}+%
\mathbf{a}}+c_{\mathbf{r}}c_{\mathbf{r}+\mathbf{a}}\right) +h.c.  \notag \\
&&+t\sum_{\mathbf{r}}\left( 2c_{\mathbf{r}}^{\dagger }c_{\mathbf{r}%
}-1\right) .
\end{eqnarray}%
Now, we consider a finite lattice system on a cylindrical geometry by taking
the periodic boundary condition in one direction and open boundary in
another direction. For a $M\times N$\ Kitaev model, the Majorana Hamiltonian
can be explicitly expressed as

\begin{eqnarray}
H_{BW} &=&-\frac{it}{2}\sum_{m=1}^{M}%
\sum_{n=1}^{N}(a_{m,n}b_{m,n}+b_{m+1,n}a_{m,n}  \notag \\
&&+b_{m,n+1}a_{m,n}-h.c.),
\end{eqnarray}%
by taking $\mathbf{r=}m\mathbf{i+}n\mathbf{j}\rightarrow (m,n)$. The
boundary conditions are $b_{m,1}=b_{m,N+1},a_{M+1,n}=0,b_{M+1,n}=0$.

Consider the Fourier transformations of Majorana operators

\begin{eqnarray}
a_{m,n} &=&\frac{1}{\sqrt{N}}\sum_{K}e^{-iKn}a_{m,K}, \\
b_{m,n} &=&\frac{1}{\sqrt{N}}\sum_{K}e^{-iKn}b_{m,K},
\end{eqnarray}%
where the wave vector $K=2\pi l/N$, $l=1,...,N$.\ Here $a_{m,K}$\ and $%
b_{m,K}$\ represent the linear combinations of\ Majorana fermion operator.
These are not standard Majorana fermions since%
\begin{equation}
a_{m,K}^{\dag }=a_{m,-K},b_{m,K}^{\dag }=b_{m,-K},
\end{equation}%
except the case with $K=0$, where%
\begin{eqnarray}
a_{m,0}^{\dag } &=&a_{m,0}=\frac{1}{\sqrt{N}}\sum_{n=1}^{N}a_{m,n}, \\
b_{m,0}^{\dag } &=&b_{m,0}=\frac{1}{\sqrt{N}}\sum_{n=1}^{N}b_{m,n},
\end{eqnarray}%
are also Majorana fermion operators. The following analysis for edge modes
only involves two such operators.

The Hamiltonian $H_{BW}$\ accordingly can be rewritten as%
\begin{eqnarray}
H_{BW} &=&\sum_{K}h_{BW}^{K}, \\
h_{BW}^{K} &=&-\frac{it}{2}\sum_{m=1}^{M}[\left( 1-e^{iK}\right)
a_{m,K}b_{m,-K}  \notag \\
&&+b_{m+1,-K}a_{m,K}-h.c.],
\end{eqnarray}%
which obeys%
\begin{equation}
\lbrack h_{BW}^{K},h_{BW}^{K^{\prime }}]=0,
\end{equation}%
i.e., $H_{BW}$ has been block diagonalized. We note that for $K=0$, we have%
\begin{equation}
h_{BW}^{0}=-\frac{it}{2}\sum_{m=1}^{M}(b_{m+1,0}a_{m,0}-h.c.).
\end{equation}%
Term $a_{m,0}b_{m,0}$ disappears from the Hamiltonian, indicating the
existence of an edge modes of Majorana fermions $a_{m,0}$\ and $b_{m,0}$. It
is a perfect edge mode with zero character decay length. The mechanism of
the mode is the fact that, a honeycomb tube lattice with zigzag boundary is
equivalent to a set of SSH chains \cite{SL}. The formation of such a state
is the result of destructive interference at the edge\textbf{.}\ Fig. \ref%
{fig1} schematically illustrates the relation among the Kitaev model on a
square lattice and the corresponding Majorana ferimonic model on a brick
wall model, and the perfect edge modes, through a small size system.

Actually, Hamiltonian $h_{BW}^{0}$ can be diagonalized by introducing $M$
fermionic operators through

\begin{equation}
d_{m}=\frac{1}{2}(a_{m,0}-ib_{m+1,0}),d_{M}=\frac{1}{2}(a_{1,0}-ib_{M,0}),
\end{equation}%
for $m=1,...,M-1$. Note operators that $d_{m}$ $(m\neq M)$ combine the
Majorana operators which derive from neighboring sites, while $d_{M}$\
combines the two ending Majorana operators. Using the above definition of\ $%
d_{m}$, the Hamiltonian $h_{BW}^{0}$\ can be written as the diagonal form

\begin{equation}
h_{BW}^{0}=2t\sum_{m=1}^{M-1}(d_{m}^{\dagger }d_{m}-\frac{1}{2})+0\times
d_{M}^{\dag }d_{M}.
\end{equation}%
On the other hand, we note that%
\begin{equation}
\lbrack d_{M},h_{BW}^{0}]=[d_{M},H_{BW}]=0,
\end{equation}%
which means that $d_{M}$\ and $d_{M}^{\dag }$\ are the eigen operators of
the Hamiltonian $H_{BW}$ with zero energy. Operators\textbf{\ }$d_{M}$%
\textbf{\ }and\textbf{\ }$d_{M}^{\dag }$\textbf{\ }are refered as
zero-energy mode operators, or edge-mode operators since only edge Majorana
fermions\textbf{\ }$a_{1,0}$ and $b_{M,0}$\textbf{\ }are involved\textbf{.}
For an arbitrary eigenstate $\left\vert \Phi \right\rangle $ of $H_{BW}$
with eigenenergy $E$, i.e.,%
\begin{equation}
H_{BW}\left\vert \Phi \right\rangle =E\left\vert \Phi \right\rangle ,
\end{equation}%
state $d_{M}\left\vert \Phi \right\rangle $\ $(d_{M}^{\dag }\left\vert \Phi
\right\rangle )$\ is also an eigenstate of $H_{BW}$\ with the same
eigenenergy $E$, if $d_{M}\left\vert \Phi \right\rangle \neq 0$ $%
(d_{M}^{\dag }\left\vert \Phi \right\rangle \neq 0)$. In general, all the
eigenstates of $H_{BW}$\ can be classified into two groups $\{\left\vert
\Phi _{+}\right\rangle \}$\ and $\{\left\vert \Phi _{-}\right\rangle \}$,
which are constructed as the forms%
\begin{equation}
\left\vert \Phi _{-}\right\rangle =\prod_{\{j\},j\neq M}d_{j}^{\dag
}\left\vert d-Vac\right\rangle ,\left\vert \Phi _{+}\right\rangle
=d_{M}^{\dag }\left\vert \Phi _{-}\right\rangle .  \label{Phi +-}
\end{equation}%
Here $\left\vert d-Vac\right\rangle $ is the normalized vacuum state of all
fermion operators $d_{j}$ $(j\in \lbrack 1,M])$%
\begin{equation}
\left\vert d-Vac\right\rangle =\Lambda \prod_{j=1}^{M-1}d_{j}\left\vert
Vac\right\rangle ,
\end{equation}%
satisfying $d_{j}\left\vert d-Vac\right\rangle =0$,\ where $\Lambda $ is the
normalization factor. Obviously we have
\begin{equation}
d_{M}\left\vert \Phi _{-}\right\rangle =0.
\end{equation}%
We find that $\left\vert \Phi _{-}\right\rangle $\ and $\left\vert \Phi
_{+}\right\rangle $ possess the same eigen energy by acting with the
commutation relation $[d_{M}^{\dag },H_{BW}]=0$ on state $\left\vert \Phi
_{-}\right\rangle $. Therefore, we conclude that all the eigenstates of $%
H_{BW}$\ is at least doubly degenerate and this degeneracy is associated
with the existence of Majorana edge modes.

\begin{figure}[tbp]
\centering
\includegraphics[bb=329 138 818 522, width=8 cm, clip]{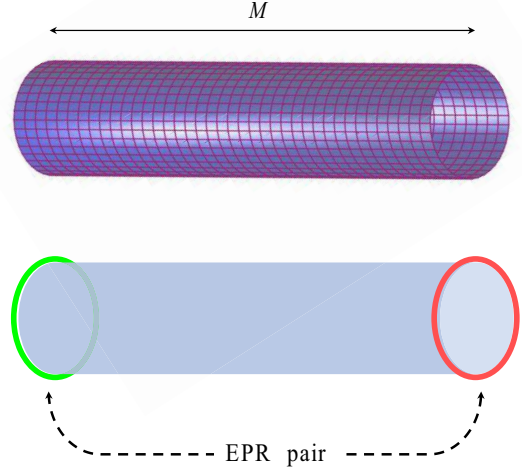}
\caption{(Color online) Schematics of the Kitaev model on a square lattice
of cylindrical geometry with length $M$ (upper panel). The Majorana
zero-mode state corresponds to an EPR pair state of spinless fermions on the
two edges of the cylinder (lower panel).}
\label{fig2}
\end{figure}
We are interested in the feature of edge-mode operator $d_{M}$. It is easy
to check that%
\begin{equation}
d_{M}=\frac{1}{2}(c_{1,0}^{\dag }+c_{1,0}-c_{M,0}^{\dag }+c_{M,0}),
\end{equation}%
where
\begin{eqnarray}
c_{1,0} &=&\frac{1}{\sqrt{N}}\sum_{n=1}^{N}c_{1,n}, \\
c_{M,0} &=&\frac{1}{\sqrt{N}}\sum_{n=1}^{N}c_{M,n},
\end{eqnarray}%
are collective fermionic operators on the two edges of the cylinder. We note
that edge-mode operator $d_{M}$\ is a linear combination of particle and
hole operators of spinless fermion $c$\ on the edge with the identical
amplitudes. To demonstrate the feature of the operator $d_{M}$, we focus on
two related states, vacuum state and excited of $d_{M}$\ particle (or hole
and particle states).\ The vacuum state of fermion operator $d_{M}$\ can be
constructed from $c$\ vacuum state as%
\begin{eqnarray}
\left\vert M-Vac\right\rangle &=&\sqrt{2}d_{M}\left\vert Vac\right\rangle
\notag \\
&=&\frac{1}{\sqrt{2}}(c_{1,0}^{\dag }-c_{M,0}^{\dag })\left\vert
Vac\right\rangle ,
\end{eqnarray}%
which satisfies $d_{M}\left\vert M-Vac\right\rangle =0$. Then the particle
state is%
\begin{equation}
d_{M}^{\dag }\left\vert M-Vac\right\rangle =\frac{1}{\sqrt{2}}%
(1+c_{M,0}^{\dag }c_{1,0}^{\dag })\left\vert Vac\right\rangle .
\end{equation}%
Remarkably, by the mappings of $\left\vert Vac\right\rangle \rightarrow
\left\vert \downarrow \right\rangle _{M}\left\vert \downarrow \right\rangle
_{1}$\ and $c_{M,0}^{\dag }c_{1,0}^{\dag }\left\vert Vac\right\rangle
\rightarrow \left\vert \uparrow \right\rangle _{M}\left\vert \uparrow
\right\rangle _{1}$, which are based on the Jordan-Wigner transformation, we
find that the edge particle state $d_{M}^{\dag }\left\vert
M-Vac\right\rangle $\ is a maximally entangled state between two edges of
the cylinder (see Fig. \ref{fig2}). On the other hand, if we take the
mapping\ $c_{1,0}^{\dag }\left\vert Vac\right\rangle \rightarrow \left\vert
\uparrow \right\rangle _{1}\left\vert \downarrow \right\rangle _{M}$\ and\ $%
c_{M,0}^{\dag })\left\vert Vac\right\rangle \rightarrow \left\vert
\downarrow \right\rangle _{1}\left\vert \uparrow \right\rangle _{M}$,\ we
find that the edge hole state\ $\left\vert M-Vac\right\rangle $\ is also a
maximally entangled state.\ Although both states $\left\vert
M-Vac\right\rangle $\ and\ $d_{M}^{\dag }\left\vert M-Vac\right\rangle $\
are not eigenstates of\ $H_{BW}$, the entanglement reflects the feature of
the edge modes.

In short, a zero-energy mode is characterized by a conventional fermion
operator, which is also referred as edge-mode operator. Any standard fermion
operator has its own vacuum and particle states, or hole and particle
states. We have shown that the corresponding hole and particle states for
the edge-mode operators\textbf{\ }$d_{M}$\textbf{\ }and\textbf{\ }$%
d_{M}^{\dag }$\textbf{\ }are both EPR pair states in the spin
representation. It reveals the non-locality of edge mode, through the
particle and hole states are not eigenstates of the system.

For the eigenstate, the long-range correlation still exists. In the section
Method, we show that the eigenstate $\left\vert \Phi _{\pm }\right\rangle $\
can be regarded as entangled states between boson and fermion. It is
expected that such a framework can be applied to more general cases.

\section{Summary}

\label{Summary} In this paper we have studied the edge modes of a finite
size Kitaev model on a square lattice. The advantage of studying the finite
system is that the obtained result can be demonstrated in synthetic lattice
system. We studied the Majorana edge modes for the Kitaev model in a
cylindrical geometry. The Majorana representation of the Hamiltonian turns
out to be equivalent to a brick wall model under some conditions. The
analytical solutions show that there exist perfect Majorana edge modes,
which are in the strong\ localization limit. We provide a new way to analyze
the excitation mechanisms in the framework of pseudospins for the edge
modes. These modes, in contrast to the modes in Kitaev chain, can appear in
small finite systems. This may provide a new venue for observing Majorana
fermions in experiments.

\section{Method}

\subsection{Pseudospin description}

\label{Pseudospin description} To get insight into the feature of the
edge-mode related eigenstates\ in such a cylindrical Kitaev model,\ we
introduce two types of pseudospin operators%
\begin{equation}
\left\{
\begin{array}{l}
s^{x}=\frac{1}{2}(c_{1,0}^{\dag }c_{M,0}+c_{M,0}^{\dag }c_{1,0}) \\
s^{y}=\frac{1}{2i}\left( c_{1,0}^{\dag }c_{M,0}-c_{M,0}^{\dag }c_{1,0}\right)
\\
s^{z}=\frac{1}{2}\left( c_{1,0}^{\dag }c_{1,0}-c_{M,0}^{\dag }c_{M,0}\right)%
\end{array}%
\right. ,
\end{equation}%
and%
\begin{equation}
\left\{
\begin{array}{l}
\tau ^{x}=\frac{1}{2}\left( c_{M,0}^{\dag }c_{1,0}^{\dag
}+c_{1,0}c_{M,0}\right) \\
\tau ^{y}=\frac{1}{2i}\left( c_{M,0}^{\dag }c_{1,0}^{\dag
}-c_{1,0}c_{M,0}\right) \\
\tau ^{z}=\frac{1}{2}\left( c_{M,0}^{\dag }c_{M,0}+c_{1,0}^{\dag
}c_{1,0}-1\right)%
\end{array}%
\right. ,
\end{equation}%
which satisfy the relations

\begin{equation}
\left[ s^{\alpha },s^{\beta }\right] =i\epsilon _{\alpha \beta \gamma
}s^{\gamma },\left[ \tau ^{\alpha },\tau ^{\beta }\right] =i\epsilon
_{\alpha \beta \gamma }\tau ^{\gamma },
\end{equation}%
and%
\begin{equation}
\left[ s^{\alpha },\tau ^{\beta }\right] =0,
\end{equation}%
where $\alpha ,\beta ,\gamma =x,y,z$. Based on these relations the
combination operator $J^{\alpha }=s^{\alpha }+\tau ^{\alpha }$ obeys the
standard angular momentum relation%
\begin{equation}
\left[ J^{\alpha },J^{\beta }\right] =i\epsilon _{\alpha \beta \gamma
}J^{\gamma }.
\end{equation}%
We note that $J^{x}$\ is a conserved observable since\ $[J^{x},H_{BW}]=0$.

On the other hand, both operators $J^{x}$ and $H_{BW}$\ are also invariant
under a local particle-hole transformation $\mathcal{P}$, which is defined as

\begin{equation}
\mathcal{P}^{-1}c_{M,0}\mathcal{P}=c_{M,0}^{\dagger },\mathcal{P}%
^{-1}c_{M-1,0}\mathcal{P}=-c_{M-1,0}^{\dagger }.  \label{P}
\end{equation}%
The fact that, $[J^{x},H_{BW}]=[\mathcal{P},H_{BW}]=[J^{x},\mathcal{P}]=0$,\
tells us operators $J^{x}$, $\mathcal{P}$ and $H_{BW}$\ share a common eigen
vectors $\left\vert \Phi _{\pm }\right\rangle $. In fact, one pseudospin can
be transformed to the other (and vice versa) by applying the transformation $%
\mathcal{P}$.\ Direct derivation shows that

\begin{equation}
\mathcal{P}^{-1}s^{2}\mathcal{P}=\tau ^{2},
\end{equation}%
and%
\begin{equation}
s^{2}+\tau ^{2}=3/4,
\end{equation}%
which result in%
\begin{equation}
\left\langle \Phi _{\pm }\right\vert s^{2}\left\vert \Phi _{\pm
}\right\rangle =\left\langle \Phi _{\pm }\right\vert \tau ^{2}\left\vert
\Phi _{\pm }\right\rangle =3/8.
\end{equation}%
Together with%
\begin{equation}
J^{x}\left\vert \Phi _{\pm }\right\rangle =\pm \frac{1}{2}\left\vert \Phi
_{\pm }\right\rangle ,
\end{equation}%
we find that state $\left\vert \Phi _{\pm }\right\rangle $\ does not have
definite values of $s$\ and $\tau $. Unlike a standard spin operator which
has its own vector space, operators $\left\{ s^{\alpha }\right\} $\ and $%
\left\{ \tau ^{\beta }\right\} $\ share a common vector space.

Actually, for two edge sub-system, there are total four\ possible states
which can be written down as%
\begin{eqnarray}
\left\vert 1\right\rangle &=&\left\vert Vac\right\rangle ,\left\vert
2\right\rangle =c_{1,0}^{\dag }\left\vert Vac\right\rangle ,  \notag \\
\left\vert 3\right\rangle &=&c_{M,0}^{\dag }\left\vert Vac\right\rangle
,\left\vert 4\right\rangle =c_{1,0}^{\dag }c_{M,0}^{\dag }\left\vert
Vac\right\rangle .
\end{eqnarray}%
We have the relations%
\begin{eqnarray}
s^{z}\left\vert 1\right\rangle &=&s^{z}\left\vert 4\right\rangle =0,  \notag
\\
s^{2}\left\vert 1\right\rangle &=&s^{2}\left\vert 4\right\rangle =0,
\label{1}
\end{eqnarray}%
and%
\begin{eqnarray}
s^{z}\left\vert 2\right\rangle &=&\frac{1}{2}\left\vert 2\right\rangle
,s^{z}\left\vert 3\right\rangle =-\frac{1}{2}\left\vert 3\right\rangle ,
\notag \\
s^{2}\left\vert 2\right\rangle &=&\frac{3}{4}\left\vert 2\right\rangle
,s^{2}\left\vert 3\right\rangle =\frac{3}{4}\left\vert 3\right\rangle ,
\label{2}
\end{eqnarray}%
which mean that states $\left\vert 1\right\rangle $\textbf{\ }and\textbf{\ }$%
\left\vert 4\right\rangle $\textbf{\ }are spin state with\textbf{\ }$s=0$%
\textbf{, }while $\left\vert 2\right\rangle $\ and $\left\vert
3\right\rangle $\ are spin states with\textbf{\ }$s=1/2$\textbf{.} Simlarly,
as for operator $\tau $, we have%
\begin{eqnarray}
\tau ^{z}\left\vert 2\right\rangle &=&\tau ^{z}\left\vert 3\right\rangle =0,
\notag \\
\tau ^{2}\left\vert 2\right\rangle &=&\tau ^{2}\left\vert 3\right\rangle =0,
\label{3}
\end{eqnarray}%
and%
\begin{eqnarray}
\tau ^{z}\left\vert 4\right\rangle &=&\frac{1}{2}\left\vert 4\right\rangle
,\tau ^{z}\left\vert 1\right\rangle =-\frac{1}{2}\left\vert 1\right\rangle ,
\notag \\
\tau ^{2}\left\vert 4\right\rangle &=&\frac{3}{4}\left\vert 4\right\rangle
,\tau ^{2}\left\vert 1\right\rangle =\frac{3}{4}\left\vert 1\right\rangle ,
\label{4}
\end{eqnarray}%
which mean that states $\left\vert 2\right\rangle $\ and $\left\vert
3\right\rangle $\ are spin state with\textbf{\ }$\tau =0$\textbf{, }while $%
\left\vert 1\right\rangle $\ and $\left\vert 4\right\rangle $\ are spin
state with\textbf{\ }$\tau =1/2$\textbf{.} Then if we regard operators $%
\left\{ s^{\alpha }\right\} $\ and $\left\{ \tau ^{\beta }\right\} $\ as
independent standard spin operators with $s$, $\tau =0,1/2$, two degree of
freedom in states $\left\vert 1\right\rangle $, $\left\vert 2\right\rangle $%
, $\left\vert 3\right\rangle $, and $\left\vert 4\right\rangle $ can be
separated and written as direct product of two independent spin states%
\begin{eqnarray}
\left\vert 1\right\rangle &=&\left\vert 0\right\rangle _{s}\left\vert
\downarrow \right\rangle _{\tau },\left\vert 4\right\rangle =\left\vert
0\right\rangle _{s}\left\vert \uparrow \right\rangle _{\tau },  \notag \\
\left\vert 2\right\rangle &=&\left\vert \uparrow \right\rangle
_{s}\left\vert 0\right\rangle _{\tau },\left\vert 3\right\rangle =\left\vert
\downarrow \right\rangle _{s}\left\vert 0\right\rangle _{\tau },
\end{eqnarray}%
where%
\begin{eqnarray}
s^{\alpha }\left\vert 0\right\rangle _{s} &=&0,  \notag \\
s^{z}\left\vert \uparrow \right\rangle _{s} &=&\frac{1}{2}\left\vert
\uparrow \right\rangle _{s},s^{z}\left\vert \downarrow \right\rangle _{s}=-%
\frac{1}{2}\left\vert \downarrow \right\rangle _{s},
\end{eqnarray}%
and%
\begin{eqnarray}
\tau ^{\beta }\left\vert 0\right\rangle _{\tau } &=&0,  \notag \\
\tau ^{z}\left\vert \uparrow \right\rangle _{\tau } &=&\frac{1}{2}\left\vert
\uparrow \right\rangle _{\tau },\tau ^{z}\left\vert \downarrow \right\rangle
_{\tau }=-\frac{1}{2}\left\vert \downarrow \right\rangle _{\tau }.
\end{eqnarray}%
Obviously, this factorization of states is consistent with the Eqs. from \ref%
{1}\ to \ref{4}. In the spirit of this representation, one can construct
equivalent states $\left\vert \widetilde{\Phi }_{\pm }\right\rangle $\ to $%
\left\vert \Phi _{\pm }\right\rangle $\ by regarding operators $\left\{
s^{\alpha }\right\} $\ and $\left\{ \tau ^{\beta }\right\} $\ as standard
spin operators with $s$, $\tau =0,1/2$,%
\begin{eqnarray}
\left\vert \widetilde{\Phi }_{+}\right\rangle &=&\frac{1}{\sqrt{2}}%
(\left\vert 0\right\rangle _{s}\left\vert \rightarrow \right\rangle _{\tau }%
\mathbf{+}\left\vert \rightarrow \right\rangle _{s}\left\vert 0\right\rangle
_{\tau }), \\
\left\vert \widetilde{\Phi }_{-}\right\rangle &=&\frac{1}{\sqrt{2}}%
(\left\vert 0\right\rangle _{s}\left\vert \leftarrow \right\rangle _{\tau }%
\mathbf{+}\left\vert \leftarrow \right\rangle _{s}\left\vert 0\right\rangle
_{\tau }),
\end{eqnarray}%
where%
\begin{eqnarray}
s^{x}\left\vert \rightarrow \right\rangle _{s} &=&\frac{1}{2}\left\vert
\rightarrow \right\rangle _{s},s^{x}\left\vert \leftarrow \right\rangle
_{s}=-\frac{1}{2}\left\vert \leftarrow \right\rangle _{s}, \\
\tau ^{x}\left\vert \rightarrow \right\rangle _{\tau } &=&\frac{1}{2}%
\left\vert \rightarrow \right\rangle _{\tau },\tau ^{x}\left\vert \leftarrow
\right\rangle _{\tau }=-\frac{1}{2}\left\vert \leftarrow \right\rangle
_{\tau }.
\end{eqnarray}%
We find that $\left\vert \widetilde{\Phi }_{\pm }\right\rangle $\ has the
same feature with $\left\vert \Phi _{\pm }\right\rangle $, i.e.,%
\begin{eqnarray}
\left\langle \widetilde{\Phi }_{\pm }\right\vert s^{2}\left\vert \widetilde{%
\Phi }_{\pm }\right\rangle &=&\left\langle \widetilde{\Phi }_{\pm
}\right\vert \tau ^{2}\left\vert \widetilde{\Phi }_{\pm }\right\rangle =3/8,
\\
J^{x}\left\vert \widetilde{\Phi }_{\pm }\right\rangle &=&\pm \frac{1}{2}%
\left\vert \widetilde{\Phi }_{\pm }\right\rangle .
\end{eqnarray}%
It indicates that eigenstate state $\left\vert \Phi _{\pm }\right\rangle $
originates from the couple of two types of excitations, boson and fermion.
Particles $s$ and $\tau $\ have an internal degree of freedom, with quantum
number $0$\ and $1/2$, corresponding to bosonic and fermionic states. State $%
\left\vert \Phi _{\pm }\right\rangle $ can be regarded as the eigenstate of
a boson-fermion coupling system. The state is maximally entangled between
particles $s$ and $\tau $\ with the respect to the boson and fermion modes.
Such an exotic feature is responsible to the existence of edge modes.

\section*{Acknowledgements}

We acknowledge the support of the CNSF (Grant No. 11374163).

\section*{Author Contributions}

Z.S. conceived the idea and carried out the study. P.W., S.L. and G.Z.
discussed the results. Z.S. wrote the manuscript with inputs from all the
other authors.

\section*{Additional information}

\textbf{Competing financial interests:} \textbf{The authors declare no
competing interests}.


\begin{thebibliography}{99}
\bibitem{Hasan} Hasan, M. Z. \& Kane, C. L. Colloquium: Topological
insulators. \textit{Rev. Mod. Phys.} \textbf{82,} 3045 (2010).

\bibitem{XLQ} Qi, X. L. \& Zhang, S. C. Topological insulators and
superconductors. \textit{Rev. Mod. Phys}. \textbf{83,} 1057 (2011).

\bibitem{CKC} Chiu, C. K., Teo, J. C. Y., Schnyder, A. P. \& Ryu, S.
Classification of topological quantum matter with symmetries. \textit{Rev.
Mod. Phys.} \textbf{88,} 035005 (2016).

\bibitem{HMW} Weng, H. M., Yu, R., Hu, X., Dai, X. \& Fang, Z. Quantum
anomalous Hall effect and related topological electronic states. \textit{%
Adv. Phys.} \textbf{64,} 227-282 (2015).

\bibitem{LF} Fu, L. \& Kane, C. L. Superconducting Proximity Effect and
Majorana Fermions at the Surface of a Topological Insulator. \textit{Phys.
Rev. Lett.} \textbf{100,} 096407 (2008).

\bibitem{RML} Lutchyn, R. M., Sau, J. D. \& Das. Sarma, S. Majorana Fermions
and a Topological Phase Transition in Semiconductor-Superconductor
Heterostructures. \textit{Phys. Rev. Lett}. \textbf{105,} 077001 (2010).

\bibitem{VM} Mourik, V., Zuo, K., Frolov, S. M., Plissard, S. R., Bakkers,
E. P. A. M. \& Kouwenhoven, L. P. Signatures of Majorana Fermions in Hybrid
Superconductor-Semiconductor Nanowire Devices. \textit{Science} \textbf{336,}
1003-1007 (2012).

\bibitem{SNP} Perge, S. N. et al. Observation of Majorana fermions in
ferromagnetic atomic chains on a superconductor. \textit{Science} \textbf{%
346,} 602-607 (2014).

\bibitem{YO} Oreg, Y., Refael, G. \& Von Oppen, F. Helical Liquids and
Majorana Bound States in Quantum Wires. \textit{Phys. Rev. Lett.} \textbf{%
105,} 177002 (2010).

\bibitem{NR} Read, N. \& Green, D. Paired states of fermions in two
dimensions with breaking of parity and time-reversal symmetries and the
fractional quantum Hall effect. \textit{Phys. Rev. B} \textbf{61,} 10267
(2000).

\bibitem{AHCN} Neto, A. H. C., Guinea, F., Peres, N. M. R., Novoselov, K. S.
\& Geim, A. K. The electronic properties of graphene. \textit{Rev. Mod. Phys}%
. \textbf{81,} 109 (2009).

\bibitem{ZKL} Liu, Z. K. et al. A stable three-dimensional topological Dirac
semimetal Cd$_{\text{3}}$As$_{\text{2}}$. \textit{Nat. Mater.} \textbf{13,}
677-681 (2014).

\bibitem{ZKL2} Liu, Z. K. et al. Discovery of a Three-Dimensional
Topological Dirac Semimetal, Na$_{\text{3}}$Bi. \textit{Science} \textbf{343,%
} 864-867 (2014).

\bibitem{JAS} Steinberg, J. A. et al. Bulk Dirac Points in Distorted
Spinels. \textit{Phys. Rev. Lett}. \textbf{112,} 036403 (2014).

\bibitem{ZW} Wang, Z. J. et al. Dirac semimetal and topological phase
transitions in A$_{\text{3}}$Bi (A=Na, K, Rb). \textit{Phys. Rev. B} \textbf{%
85,} 195320 (2012).

\bibitem{JX} Xiong, J. et al. Evidence for the chiral anomaly in the Dirac
semimetal Na$_{\text{3}}$Bi. \textit{Science} \textbf{350,} 413-416 (2015).

\bibitem{SMY} Young, S. M., Zaheer, S., Teo, J. C. Y., Kane, C. L., Mele, E.
J. \& Rappe, A. M. Dirac Semimetal in Three Dimensions. \textit{Phys. Rev.
Lett.} \textbf{108,} 140405 (2012).

\bibitem{MH} Hirschberger, M. et al. The chiral anomaly and thermopower of
Weyl fermions in the half-Heusler GdPtBi. \textit{Nat. Mater}. \textbf{15, }%
1161-1165 (2016).

\bibitem{SMH} Huang, S. M. et al. A Weyl Fermion semimetal with surface
Fermi arcs in the transition metal monopnictide TaAs class. \textit{Nat.
Commun}. \textbf{6,} 7373 (2015).

\bibitem{BQL} Lv, B. Q. et al. Experimental Discovery of Weyl Semimetal
TaAs. \textit{Phys. Rev. X} \textbf{5,} 031013 (2015).

\bibitem{BQL2} Lv, B. Q. et al. Observation of Weyl nodes in TaAs. \textit{%
Nat. Phys}. \textbf{11,} 724-727 (2015).

\bibitem{CS} Shekhar, C. et al. Observation of chiral magneto-transport in
RPtBi topological Heusler compounds. ArXiv:1604.01641 (2016).

\bibitem{XW} Wan, X., Turner, A. M., Vishwanath, A. \& Savrasov, S. Y.
Topological semimetal and Fermi-arc surface states in the electronic
structure of pyrochlore iridates. \textit{Phys. Rev. B \textbf{83},} 205101
(2011).

\bibitem{HW} Weng, H., Fang, C., Fang, Z., Bernevig, B. A. \& Dai, X. Weyl
Semimetal Phase in Noncentrosymmetric Transition-Metal Monophosphides.
\textit{Phys. Rev. X }\textbf{5,} 011029 (2015).

\bibitem{SYX} Xu, S. Y. et al. Discovery of a Weyl fermion state with Fermi
arcs in niobium arsenide. \textit{Nat. Phys}. \textbf{11,} 748-754 (2015).

\bibitem{SYX2} Xu, S. Y. et al. Discovery of a Weyl fermion semimetal and
topological Fermi arcs. \textit{Science} \textbf{349} 613-617 (2015).

\bibitem{AJ} Alicea, J. New directions in the pursuit of majorana fermions
in solid state systems. \textit{Rep. Prog. Phys}. \textbf{75}, 076501 (2012).

\bibitem{BCWJ} Beenakker, C. W. J. Search for Majorana Fermions in
Superconductors. \textit{Annu. Rev. Condens. Matter Phys.} \textbf{4},
113-136 (2013).

\bibitem{STD} Stanescu, T. D. \& Tewari, S. Majorana fermions in
semiconductor nanowires: Fundamentals, modeling, and experiment.\textit{\ J.
Phys. Condens. Matter} \textbf{25}, 233201 (2013).

\bibitem{LM} Leijnse, M. \& Flensberg, K. Introduction to topological
superconductivity and majorana fermions. \textit{Semicond. Sci. Technol.}
\textbf{27}, 124003 (2012).

\bibitem{ESR} Elliott, S. R. \& Franz, M. Colloquium: Majorana fermions in
nuclear, particle, and solid-state physics. \textit{Rev. Mod. Phys.} \textbf{%
87}, 137 (2015).

\bibitem{DSS} Das Sarma, S., Freedman, M. \& Nayak, C. Majorana zero modes
and topological quantum computation. \textit{NPJ Quantum Information}
\textbf{1}, 15001 (2015).

\bibitem{SM} Sato, M. \& Fujimoto, S. Majorana fermions and topology in
superconductors. \textit{J. Phys. Soc. Jpn}. \textbf{85}, 072001 (2016).

\bibitem{NC} Nayak, C., Simon, S. H., Stern, A., Freedman, M. \& Das Sarma,
S. Non-abelian anyons and topological quantum computation. \textit{Rev. Mod.
Phys.} \textbf{80}, 1083 (2008).

\bibitem{SJD} Sau, J. D., Lutchyn, R. M., Tewari, S. \& Das Sarma, S.
Generic New Platform for Topological Quantum Computation Using Semiconductor
Heterostructures. \textit{Phys. Rev. Lett. }\textbf{104}, 040502 (2010).

\bibitem{AJ2} Alicea, J. Majorana fermions in a tunable semiconductor
device. \textit{Phys. Rev. B }\textbf{81}, 125318 (2010).

\bibitem{QXL} Qi, X. L., Hughes, T. L., \& Zhang, S. C. Chiral topological
superconductor from the quantum hall state. \textit{Phys. Rev. B}\emph{\ }%
\textbf{82}, 184516 (2010).

\bibitem{CSB} Chung, S. B., Qi, X. L., Maciejko, J. \& S. C. Zhang.
Conductance and noise signatures of majorana backscattering. \textit{Phys.
Rev. B} \textbf{83}, 100512 (2011).

\bibitem{LKT} Law, K. T., Lee, P. A. \& Ng, T. K. Majorana fermion induced
resonant andreev reflection. \textit{Phys. Rev. Lett.} \textbf{103}, 237001
(2009).

\bibitem{AAR} Akhmerov, A. R., Nilsson, Johan \& J. Beenakker, C. W.
Electrically detected interferometry of majorana fermions in a topological
insulator. \textit{Phys. Rev. Lett}. \textbf{102}, 216404 (2009).

\bibitem{AJ3} Alicea, J., Oreg, Y., Refael, G., von Oppen, F. \& Fisher, M.
P. Non-abelian statistics and topological quantum information processing in
1d wire networks. \textit{Nat. Phys.} \textbf{7}, 412-417 (2011).

\bibitem{LRM2} Lutchyn, R. M., Stanescu, T. D. \& Das Sarma, S. Search for
Majorana Fermions in Multiband Semiconducting Nanowires. \textit{Phys. Rev.
Lett.} \textbf{106}, 127001 (2011).

\bibitem{STD2} Stanescu, T. D., Lutchyn, R. M. \& Das Sarma, S. Majorana
fermions in semiconductor nanowires. \textit{Phys. Rev. B} \textbf{84},
144522 (2011).

\bibitem{PAC} Potter, A. C. \& Lee, P. A. Multichannel Generalization of
Kitaev's Majorana End States and a Practical Route to Realize Them in Thin
Films.\textit{\ Phys. Rev. Lett.} \textbf{105}, 227003 (2010).

\bibitem{RD} Rainis, D., Trifunovic, L., Klinovaja, J. \& Loss, D. Towards a
realistic transport modeling in a superconducting nanowire with majorana
fermions. \textit{Phys. Rev. B }\textbf{87}, 024515 (2013).

\bibitem{PE} Prada, E., San Jose, P. \& Aguado, R. Transport spectroscopy of
ns nanowire junctions with majorana fermions. \textit{Phys. Rev. B} \textbf{%
86}, 180503 (2012).

\bibitem{DSS2} Das Sarma, S., Sau, J. D. \& Stanescu, T. D. Splitting of the
zero-bias conductance peak as smoking gun evidence for the existence of the
majorana mode in a superconductor-semiconductor nanowire. \textit{Phys. Rev.
B} \textbf{86}, 220506 (2012).

\bibitem{LJ} Liu, J., Song, J. T., Sun, Q. F. \& Xie X. C. Even-odd
interference effect in a topological superconducting wire. \textit{Phy Rev. B%
} \textbf{96}, 195307 (2017)\textbf{.}

\bibitem{SM2} Sato, M., Takahashi, Y. \& Fujimoto, S. Non-Abelian
Topological Order in s-Wave Superfluids of Ultracold Fermionic Atoms.
\textit{Phys. Rev. Lett}. \textbf{103}, 020401 (2009).

\bibitem{ZC} Zhang, C., Tewari, S., Lutchyn, R. M. \& Das Sarma, S. Px+iPy
Superfluid from S-Wave Interactions of Fermionic Cold Atoms. \textit{Phys.
Rev. Lett.} \textbf{101}, 160401 (2008).

\bibitem{TS} Tewari, S., Das Sarma, S., Nayak, C., Zhang, C. \& Zoller, P.
Quantum Computation using Vortices and Majorana Zero Modes of a \textit{p}$_{%
\text{x}}$\textit{+ip}$_{\text{y}}$ Superfluid of Fermionic Cold Atoms.
\textit{Phys. Rev. Lett.} \textbf{98}, 010506 (2007).

\bibitem{LXJ} Liu, X. J., Law, K. T. \& Ng, T. K. Realization of 2D
Spin-Orbit Interaction and Exotic Topological Orders in Cold Atoms. \textit{%
Phys. Rev. Lett.} \textbf{112}, 086401 (2014).

\bibitem{JL} Jiang, L. et. al. Majorana Fermions in Equilibrium and in
Driven Cold-Atom Quantum Wires. \textit{Phys. Rev. Lett. }\textbf{106},
220402 (2011).

\bibitem{DS} Diehl, S., Rico, E., Baranov, M. A. \& Zoller, P. Topology by
dissipation in atomic quantum wires. \textit{Nat. Phys. }\textbf{7}, 971-977
(2011).

\bibitem{RLP} Rokhinson, L. P., Liu, X. \& Furdyna, J. K. The fractional ac
Josephson effect in a semiconductor-superconductor nanowire as a signature
of majorana particles. \textit{Nat. Phys.} \textbf{8}, 795-799 (2012).

\bibitem{DMT} Deng, M. T., Yu, C. L., Huang, G. Y., Larsson, M., Caroff, P.
\& Xu, H. Q. Anomalous zero-bias conductance peak in a nb--insb nanowire--nb
hybrid device. \textit{Nano Lett. }\textbf{12}, 6414 (2012).

\bibitem{DA} Das, A., Ronen, Y., Most, Y., Oreg, Y., Heiblum, M. \&
Shtrikman, H. Zero-bias peaks and splitting in an Al-InAs nanowire
topological superconductor as a signature of majorana fermions. \textit{Nat.
Phys}. \textbf{8}, 887-895 (2012).

\bibitem{FADK} Finck, A. D. K., Van Harlingen, D. J., Mohseni, P. K., Jung,
K. \& Li, X. Anomalous Modulation of a Zero-Bias Peak in a Hybrid
Nanowire-Superconductor Device. \textit{Phys. Rev. Lett.} \textbf{110},
126406 (2013).

\bibitem{CHOH} Churchill, H. O. H., Fatemi, V., Grove Rasmussen, K., Deng,
M. T., Caroff, P., Xu, H. Q. \& Marcus, C. M. Superconductor-nanowire
devices from tunneling to the multichannel regime: Zero-bias oscillations
and magnetoconductance crossover. \textit{Phys. Rev. B} \textbf{87}, 241401
(2013).

\bibitem{XJP} Xu, J. P. et. al. Experimental Detection of a Majorana Mode in
the Core of a Magnetic Vortex inside a Topological Insulator-Superconductor
Bi$_{\text{2}}$Te$_{\text{3}}$/NbSe$_{\text{2}}$ Heterostructure. \textit{%
Phys. Rev. Lett}. \textbf{114}, 017001 (2015).

\bibitem{SHH} Sun, H. H. et. al. Majorana Zero Mode Detected with Spin
Selective Andreev Reflection in the Vortex of a Topological Superconductor.
\textit{Phys. Rev. Lett}. \textbf{116}, 257003 (2016).

\bibitem{WMX} Wang, M. X. et al. The coexistence of superconductivity and
topological order in the Bi$_{\text{2}}$Se$_{\text{3}}$ thin films. \textit{%
Science} \textbf{336}, 52-55 (2012).

\bibitem{WZF} Wang, Z. F., et. al. Topological edge states in a
high-temperature superconductor FeSe/SrTiO$_{\text{3}}$(001) film. \textit{%
Nat. Mater.} \textbf{15}, 968-973 (2016).

\bibitem{AK} Kitaev, A. Anyons in an exactly solved model and beyond.
\textit{Ann Phys}. \textbf{321,} 2-111 (2006).

\bibitem{GB} Baskaran, G., Mandal, S. \& Shankar, R. Exact Results for Spin
Dynamics and Fractionalization in the Kitaev Model. \textit{Phys. Rev. Lett}%
. \textbf{98,} 247201 (2007).

\bibitem{DHL} Lee, D. H., Zhang, G. M. \& Xiang, T. Edge Solitons of
Topological Insulators and Fractionalized Quasiparticles in Two Dimensions.
\textit{Phys. Rev. Lett}. \textbf{99,} 196805 (2007).

\bibitem{KPS} Schmidt, K. P., Dusuel, S. \& Vidal, J. Emergent Fermions and
Anyons in the Kitaev Model. \textit{Phys. Rev. Lett.} \textbf{100,} 057208
(2008).

\bibitem{GK} Kells, G. et al. Topological Degeneracy and Vortex Manipulation
in Kitaev's Honeycomb Model. Phys \textit{Rev. Lett}. \textbf{101,} 240404
(2008).

\bibitem{GK2} Kells, G. Slingerland, J. K. \& Vala, J. Description of
Kitaev's honeycomb model with toric-code stabilizers. \textit{Phys. Rev. B }%
\textbf{80,} 125415 (2009).

\bibitem{GK3} Kells, G. \& Vala, J. Zero energy and chiral edge modes in a
p-wave magnetic spin model. \textit{Phys. Rev. B} \textbf{82,} 125122 (2010).

\bibitem{AYK} Kitaev, A. Y. Unpaired Majorana fermions in quantum wires.
\textit{Physics-Uspekhi} \textbf{44,} 131 (2001).

\bibitem{LPR} Rokhinson, L. P., Liu, X. \& Furdyna, J. K. The fractional
a.c. Josephson effect in a semiconductor--superconductor nanowire as a
signature of Majorana particles. \textit{Nat. Phys}. \textbf{8}, 795-799
(2012).

\bibitem{BA} Banerjee, A. et al. Proximate Kitaev quantum spin liquid
behaviour in a honeycomb magnet. \textit{Nature Materials}. \textbf{15,}
733-740 (2016).

\bibitem{SL} Lin, S., Zhang, G., Li, C. \& Song, Z. Magnetic-flux-driven
topological quantum phase transition and manipulation of perfect edge states
in graphene tube. \textit{Sci. Rep}. \textbf{6,} 31953 (2016); G. Zhang, C.
Li, and Z. Song. Majorana charges, winding numbers and Chern numbers in
quantum Ising models. \textit{Sci. Rep}. \textbf{7,} 8176 (2017).
\end{thebibliography}
\end{document}